\documentclass[10pt]{article}

\usepackage{amsmath}
\usepackage{amssymb}
\usepackage{courier}

\usepackage{graphicx}

\usepackage{cite}

\usepackage{color} 

\usepackage[labelfont=bf,labelsep=period,justification=raggedright]{caption}

\makeatletter
\renewcommand{\@biblabel}[1]{\quad#1.}
\makeatother

\definecolor{changecolour}{rgb}{0,0,0}

\date{}

\begin{document}

\begin{flushleft}
{\LARGE \noindent Targeted social mobilization in a global manhunt}\\
\vspace{0.1cm}
Alex Rutherford$^{\hspace{0.1cm}1,*}$,
Manuel Cebrian$^{\hspace{0.1cm}3,4,*}$,
Iyad Rahwan$^{\hspace{0.1cm}1,2,*,\dag}$,
Sohan Dsouza$^{\hspace{0.1cm}1}$, 
James McInerney$^{\hspace{0.1cm}5}$, 
Victor Naroditskiy$^{\hspace{0.1cm}5}$, 
Matteo Venanzi$^{\hspace{0.1cm}5}$, 
Nicholas R. Jennings$^{\hspace{0.1cm}5}$, 
J.R. deLara$^{\hspace{0.1cm}6}$, 
Eero Wahlstedt$^{\hspace{0.1cm}7}$,
Steven U. Miller$^{\hspace{0.1cm}8}$
\\
\vspace{0.1cm}
\bf\scriptsize{$^1$} 
$^{1}$Computing and Information Science, Masdar Institute of Science and Technology, Abu Dhabi, UAE; 
$^{2}$School of Informatics, University of Edinburgh, Edinburgh, UK; 
$^{3}$National Information and Communications Technology Australia, Melbourne, Victoria, Australia; 
$^{4}$Department of Computer Science and Engineering, University of California at San Diego, La Jolla, CA, USA; 
$^{5}$School of Electronics and Computer Science, University of Southampton, Southampton, UK;
$^{6}$George Washington University, Washington DC, USA; 
$^{7}$University of Oxford, Oxford, UK; 
$^{8}$Champlain College, Burlington, VT, USA\\
$^{*}$A.R., M.C., and I.R. contributed equally to this work.\\
$^{\dag}$To whom correspondence should be addressed; Email: irahwan@acm.org
\end{flushleft}
\noindent

\section*{Abstract}

Social mobilization, the ability to mobilize large numbers of people via social networks to achieve highly distributed tasks, has received significant attention in recent times. This growing capability, facilitated by modern communication technology, is highly relevant to endeavors which require the search for individuals that possess rare information or skills, such as finding medical doctors during disasters, or searching for missing people. An open question remains, as to whether in time-critical situations, people are able to recruit in a targeted manner, or whether they resort to so-called blind search, recruiting as many acquaintances as possible via broadcast communication. To explore this question, we examine data from our recent success in the U.S. State Department's Tag Challenge, which required locating and photographing 5 target persons in 5 different cities in the United States and Europe -- in under 12 hours -- based only on a single mug-shot. We find that people are able to consistently route information in a targeted fashion even under increasing time pressure. We derive an analytical model for social-media fueled global mobilization and use it to quantify the extent to which people were targeting their peers during recruitment. Our model estimates that approximately 1 in 3 messages were of targeted fashion during the most time-sensitive period of the challenge. %
This is a novel observation at such short temporal scales, and calls for opportunities for devising viral incentive schemes that provide distance or time-sensitive rewards to approach the target geography more rapidly. This observation of `12 hours of separation' between individuals has applications in multiple areas from emergency preparedness, to political mobilization.
\section*{Introduction}

The Internet and online social media are now credited with the unprecedented ability to coordinate the mobilization of large masses of people to achieve remarkable feats that require coverage of large geographical and informational landscapes in a very limited time. Social media has been used to mobilize volunteers to map natural disasters in real-time \cite{gao2011harnessing}, to conduct large-scale search-and-rescue missions \cite{jimgray}, and to locate physical objects within extremely short time frames \cite{balloonsScience}.

Despite the numerous successes attributed to the Internet, mobile communication and social media, we still lack a comprehensive understanding of the dynamics of technology-mediated social mobilization. Open questions remain about essential aspects that determine the success of social mobilization. One such aspect is the relationship between social interaction and geography. Social interaction is an essential driver of recruitment and coordination. However, social interaction is constrained by geography \cite{libennowelletal2005}, and such constraints exhibit fundamentally different characteristics for large communities \cite{onnela2011geographic}. Further, geography is influenced by the nature of the task at hand, as we discuss below.

Consider the task of mobilizing protesters as part of the Occupy Wall Street movement \cite{noam2012occupy}. It has recently been shown that social interaction exhibits a disproportionately high degree of geographical locality, reflecting the movement's efforts to mobilize resources in their local neighborhoods and cities \cite{conover2013geospatial,conover2013digital}.

On the other hand, mobilization for large search-and-rescue operations demands the opposite approach, namely spreading the message and recruiting participants in geographically distant locations. In the  \emph{DARPA Network Challenge} (a.k.a. \emph{Red Balloon Challenge}), organized by the Defense Advanced Research Projects Agency, teams competed to locate and submit the coordinates of 10 tethered weather balloons dispersed at random locations all over the continental United States. The winning team, based at MIT, won the challenge by locating all balloons in less than 9 hours \cite{tang2011reflecting}. The team used an incentive scheme to kick start an information and recruitment cascade that resulted in 4,400 sign-ups to the team's Web site within 48 hours. Our earlier analysis revealed that the recursive incentive scheme may have played an important role in maximizing the speed and branching of the diffusion to limits above what is normally observed in viral propagation schemes \cite{iribarrenmoro2009}. Further, data reveals that people managed to recruit acquaintances who are more distant than expected, thus contributing to the rapid coverage of a large geographical area \cite{balloonsScience}.

Another class of mobilization tasks requires geographical propagation that simultaneously spans large distances, while exhibiting targeted spatial dynamics. An example of this is search for a missing person or an object with a known approximate location. Milgram's landmark ``small world'' experiment showed that people are, in principle, able to find a target individual using 6 hops on the global social network \cite{milgram1967}. This result has been reaffirmed in the Internet age in an email-based version of Milgram's experiment \cite{doddsetal2003}. This phenomenon relies on people's ability to form reliable estimates of distance to the target, in order to exploit the large jumps afforded by small world networks as they forward the message to their acquaintances \cite{kleinberg2000,kleinberg2000stoc,adamic:adar:2005}. In particular, people rely on \emph{heuristic} information (simple rules of thumb for guiding choice) in the routing of information by the recruitment of acquaintances. Geographical distance, along with non-geographical distance measures -- such as similarity of occupation to the target individual -- form particularly effective heuristics \cite{watts2002identity}. For example, if the target is known to be a Professor residing in Kyoto, Japan, one might send it to a friend who lives in Tokyo, Japan, as they are more likely to know someone who lives in Kyoto, who in turn may know someone in academia, and so on.

An open question remains as to whether in \emph{time-critical} situations, such as public response to natural disasters, an abduction, or search for a missing child, people are still able to spread information in such a heuristic manner. Humans have a limited amount of time per day to dedicate to social interaction \cite{miritello2013time}, which poses a limit on the effort one can invest in persuading an acquaintance to act. Further, time pressure can affect the way in which people process environmental information \cite{ozel2001time}. Consequently, people may be expected to resort to so-called \emph{blind} search, focusing simply on the recruitment of as many acquaintances as possible via broadcast messaging~\cite{national2013Public}. However, while this strategy may be effective at delivering the message to a broad audience, it results in lower effort in finding and mobilizing those recruits that have high \emph{affinity} with the task (due to their location or other characteristics), and are therefore more likely to propagate the message or participate in the required action \cite{affinitymoro}.


We examined the spatial dynamics of global recruitment in the State Department's Tag Challenge, which required competing teams to locate and photograph 5 target ``thieves'' (actors) in 5 different cities in the US and Europe, based only on a mug shot released at 8:00am local time in each respective city~\cite{TagPressRelease}. The targets were only visible for 12 hours, and followed pre-arranged itineraries around the cities of Stockholm, London, Bratislava, New York City and
Washington D.C. Our team successfully located 3 of the 5 suspects \cite{TagNewScientist}, winning the competition by remotely mobilizing volunteers through social media using a recursive incentive mechanism that encourages recruitment~\cite{kleinberg2005query,STOC}. This was achieved despite the fact that none of our team members were based in any of the target cities \cite{TagIEEE}.

The challenge provided a rare opportunity to quantify the dynamics of large-scale, global social mobilization in a time-critical scenario from a spatial and temporal perspective. The 12 hour deadline provided clear urgency. Furthermore, the announcement of the challenge, 2 months in advance, provided a chance to quantify the growth of awareness over time, as we approach the actual day of the challenge, March 31st, 2012. Finally, due to its geographical dispersal over multiple countries and languages, no single small team of acquaintances can conceivably achieve the task without the help of others not directly connected to them. Consequently, people were required to forward messages to acquaintances who are either in the target cities, or whom they believed would be more likely to forward messages towards those cities. Despite the DARPA Network Challenge being very close in aim, it did not provide this opportunity, as there was no information available to the searchers whatsoever about the location of the balloons. 

We collected data about the general awareness of the challenge which is not specific to the efforts of a particular team, measured by number of hits to the main challenge organizers' Web site, as well as on major social media sites (Twitter and Facebook). We also captured data about the winning team's presence on major social media sites (Twitter and Facebook). This provides a quantitative view of the growth dynamics of mobilization over time as the deadline approaches. More importantly, by mapping the approximate geographical locations of different social media messages, we were able to quantify the geographical convergence towards the target cities.

Twitter, the popular micro-blogging service, is an ideal barometer for investigating blind versus heuristic (targeted) mobilization strategies as both modes of communication are available. Users may tweet messages to all their friends (the content is also publicly available if the user chooses this option). Alternatively, a user may mention one or more other users specifically, regardless of whether they are friends or not, by adding the symbol \texttt{@} followed by the target user's Twitter name. For example, to target a person with user name \texttt{alex}, one simply includes the string \texttt{@alex} in the message. If a tweet is of this second variety, the mentioned user receives a specific alert and is generally obliged to respond, or at least pay more attention to the message. Often, such targeted messaging also leads to subsequent public or private conversations. In the case of the Tag Challenge, such conversations can be seen as an effort exerted on behalf of the recruiter to persuade the recruit to join the cause. Although some tweets can be considered to be specific to a particular team e.g. \texttt{`Snap a picture of a traveler in this digital scavenger hunt and you could win \$ for you/charity. @TagTeam\_'} as opposed to general tweets simply raising awareness e.g. \texttt{` http://t.co/HJImNoN0 5 thieves, 5 cities, 12 hours: Can Twitter catch them? - http://t.co/sqvUBI7F'} Here we aggregate all tag-related tweets together, so we may consider our findings to be general between teams.

By classifying each challenge-related message to either the broadcast or targeted variety, we were able to investigate the extent of conscious effort towards targeted mobilization over time as the deadline approaches. In addition, by combining this information with the approximate geographical location of the target audience, it was also possible to investigate whether this targeting was effective in converging towards the target cities geographically.

It is important to disentangle two potential explanations of the phenomenon of targeted recruitment in this time-critical social mobilization. One explanation is the explicit effort on behalf of participants to identify and recruit acquaintances who are closer to the target geography. But another explanation is also possible, namely the intrinsic structure of global communication and its role in routing information automatically towards hubs. This is particularly relevant, since two of the target cities, London and New York City, are recognized global hubs, with disproportionately higher social, financial, and social media ties to the rest of the world. To disentangle the roles played by global communication structure and by individual participant choices, we developed a biased routing model that parameterizes the degree of explicit heuristic targeting, and use it to quantify the behavior observed.

\section*{Results}

\subsection*{Media Exposure}

Figure 1 shows the daily volume of Tweets related to the Tag Challenge and traffic to the official website (see Materials \& Methods).  The dates of major media articles concerning the challenge are also indicated. There is clearly some degree of correlation between media coverage and social media traffic. However significant traffic persists on days with no media coverage suggesting that there is also a slower process of peer-to-peer sharing of information about the challenge.

We also see from Figure 2 that our team's social media presence, measured by the daily number of impressions of our presence on Facebook, provided access to daily volumes of several thousand potential searchers. Although this measure counts repeated exposure by the same users, the total sums to over 29,000. The official Tag Challenge Facebook page also created over 86,000 impressions. We can therefore infer the presence of a hidden network of `passive recruits' -- people who are aware of the challenge, yet are not sufficiently motivated to sign up and recruit others, but who will report sightings of the target. Such a mechanism was found to be a necessary condition for successful social mobilisation in geographical search~\cite{rutherfordpnas}.

\subsection*{Evidence of Targeted Mobilization}

Figure 3 shows the distance scaling behaviour of traffic to the Tag Challenge Web site in the $50$ days leading up to the challenge. The distance from the originating Internet Protocol (IP) address to the nearest Tag Challenge city was calculated for each unique visitor. After filtering distance independent traffic and smoothing (see Materials \& Methods), we observe a strong trend of geographical convergence towards the target cities over time, quantified by the Pearson coefficient $(r,p)=(-0.61, <10^{-5})$. 

Figure 4 considers the \emph{rate} at which individual users are specifically targeted (i.e. \texttt{@}-mentioned) in the Tweets related to the Tag Challenge. This distinguishes messages which \emph{broadcast} to all followers from those which target \emph{specific} users perceived to be useful for locating the targets (we exclude Tweets from the participating teams from this analysis). The proportion of Twitter traffic targeting individuals increases in the 6 days leading up to the Tag Challenge $(r,p)=(0.825,0.012)$.

This trend is additionally supported by Figure 5, which considers the \emph{location} of users specifically targeted (\texttt{@}-mentioned) in Tweets. The effect of spurious noise was mitigated with the use of a 4 day moving average. The daily proportion of these targeted users located in the tag cities (defined as 25km from the city centre), with respect to the \emph{total} number of daily targeted users, was seen to increase approaching the challenge day. A strong correlation with time was found $(r,p)=(0.912,0.002)$ ($(r,p)=(0.822,0.012)$ using the raw, unsmoothed data). This result suggests that Twitter users successfully routed information geographically towards users more likely to locate a target.

The increase in both the \emph{rate} of targeted messaging and its \emph{geographical convergence} suggests that, as time becomes more critical, people become surprisingly \emph{more} rather than \emph{less} targeted in their social mobilization heuristic. This is a novel observation at such short temporal scales (days to hours), and calls for devising viral incentive schemes that provide distance- or time-sensitive rewards to approach the target geography more rapidly, with applications in multiple areas from emergency preparedness~\cite{gao2011harnessing,national2013Public} to political mobilization~\cite{gonzalez2011dynamics,bond201261}.

\textcolor{changecolour}{At this point, we emphasise that these results are not team-specific. All tweets related to the challenge were collected and analysed together. Each of
these tweets might refer to a particular team or may simply wish to draw attention to the
challenge. While we present the exposure of our team’s Facebook page in Figure 2, we also compare it to the official page of the challenge itself which may be considered
‘team-agnostic’. Figure 3 presents the geographical convergence of traffic to the official Tag
Challenge website, this traffic cannot be attributed to any particular team. 
}

\subsection*{Disentangling Targeting Behavior}

The results above suggest the existence of a significant effort by people to mobilize others in a targeted manner, moving towards the target cities. However, it is reasonable to suspect that this observed behavior may be, at least in part, an artefact of the importance of major cities like New York and London --- which may receive a disproportionately large amount of traffic regardless of the propagation process. Thus it is important to quantify the extent to which we can expect to reach those cities without any deliberate targeting (top hubs are listed in table S1 and table S2), then use this baseline to quantify the amount of targeting needed to produce the observed behavior in the Tag Challenge.

To investigate this issue, we construct a network of communications between global Metropolitan Statistical Areas (MSA). We use flight frequency data between MSAs as a proxy for social media communication intensity, which have been shown to correlate well (and more strongly than distance) with traffic from Twitter data \cite{geographytwitter}. Air traffic connections reflect the cultural/linguistic and even post-colonial and post-Commonwealth expatriate ties that have been found to be present in social networks~\cite{meshcivilisations,uganderetal2011} as well as inter-city economic relations~\cite{beaverstock} and internet connectivity~\cite{cityInternet}. The raw flight numbers and proportions of flights between cities are represented in figures S1 and S2. An additional advantage of using the air flight network is that we are able to capture the structure of what is a combination of different social media platforms which make up a fragmented global social media ecosystem. This includes not only email but also Facebook, Orkut and Weibo which dominate in North America and Europe, the Lusosphere and China respectively along with many others.

We simulate a random walk over the MSA network, which represents the diffusion of social mobilization using social media and other means of communication (see Materials and Methods for more details; the list of cities can be found in Tables S3-7). To capture the effect of different mixing of \emph{targeted} and \emph{broadcasting} behaviour, we assign some degree of geographical greediness (targeting) $g \in [0,1]$ in making the mobilization decisions. Such a random walk does not attempt to replicate the \textit{dynamics} of information flows over time. Rather it seeks to determine the static centrality of specific nodes indirectly from the stationary occupation probability distribution of a walker through resampling. The level of greediness is fixed in each simulation, and given that the degree of targeting increases as the challenge approaches, each simulation measures this property of the network at a fixed moment in time.

With probability $(1-g)$ a random walker on a node chooses to move (i.e. send a message) to a connected node randomly according to the outgoing edge weights (including self-edges capturing local communication within the MSA). With probability $g$ the walker instead moves greedily to one of its neighbours which enjoys the network-constrained, closest geographic position to any Tag Challenge city (it does this independently of the edge weight). Note that this will generally lead to an \emph{overestimation} of the centralities of the Tag Challenge cities since it assumes that people can successfully leverage any link to a Tag Challenge city no matter how weak it might be. Therefore the degree of greediness (targeting) we report to reproduce our observations should be considered a lower bound. The agent takes very large steps in space only when there exist long haul flight connections. The greedy behavior represents an agent who actively chooses to leverage social ties which are perceived to be more likely to find a target due to privileged location in space \cite{milgram1967}. The distance to the nearest tag city may be considered a \textit{heuristic} which agents use to target a particular city. Note that moving to the geographically closest city may be sub-optimal, since the new, closer city may not in fact be well connected to the target city. However agents are unlikely to have perfect knowledge of the network and so the shortest path to the target city. When a walker chooses to move greedily and has more than one Tag Challenge city among its neighbours, it chooses one at random.

We perform simulations to determine the stationary probability distributions of the above random walk (10$^{\mathrm{6}}$ steps per simulation), given various degrees of greedy targeting towards Tag Challenge cities. From this stationary probability we infer the effective centralities of the different cities.

Figure 6 (red) shows the unbiased centralities without any greedy targeted mobilization. The figure highlights the existence of clear peaks at hub cities, including some tag cities themselves. This random walk, corresponding to untargeted broadcast mobilization by participants, leads to $5\%$ of traffic ending up in one of the Tag Challenge cities. While this is a significant proportion in a global network of metropolitan areas, largely driven by the centralities of London and New York, it is significantly lower than the observed proportion. In particular, as shown in Figure 5 the proportion of targeted tweets with \texttt{@}-mentions increases to $\approx 0.7$ as the deadline approaches. The proportion of those tweets that are in one of the target cities is $\approx 0.65$ (Figure 5). This means that the proportion of messages reaching the target cities is approximately $0.7 \times 0.65 \approx 0.46$, almost an order of magnitude higher than what would be expected by an unbiased, non-targeting random flow of messages.

Figure 6 (bottom left, black) highlights that a significant degree of targeting behavior, corresponding to $g=30\%$, is required to approach the approximate proportion of time spent in the Tag Challenge cities as observed in the data (Figure 7 shows the unbiased and targeted centralities on a map). In other words, people not only need to target others with personalized recruitment messages, but they also need to do so using a geographically informed heuristic at least $30\%$ of the time. Even when restricting the communication network to North America and Europe (see figure S3), to mitigate the affects of linguistic barriers, significant targeting remains necessary to reproduce the observed proportions of traffic. However the diverse originating locations of global traffic to our team's site suggests that awareness of the challenge did transcend linguistic barriers, justifying consideration of the full global network (see figure S4).

\section*{Discussion}

Sixty years ago, social psychologist Stanley Milgram redefined our notion of social distance with his landmark \emph{Six Degrees of Separation} experiment~\cite{milgram1967}, showing that we are, on average, only 6 hops of friendship away from anyone else on earth. Facebook found the degree of separation to be only 4 in their digital network~\cite{backstrom}. Endeavors like the Tag Challenge are set to redefine our conception of the temporal and spatial limits of technology-mediated social mobilization in the Internet age, showing that using a network of mobilised people, in principle, we can find any person (who is not particularly hiding) in less than 12 hours. 

\textcolor{changecolour}{
The data analysed in this paper includes public tweets, data collected via our own team's efforts and data aceesible to us due to our privileged status as eventual winners. However we can confidently conclude that this relatively small sample is representative of the much larger flow of private communications. The conclusions drawn are not specific to our team
and since we do not have access to data collected by other teams for comparison we neccesarily limit
our analysis to this.}

We have shown that this \emph{12 hours of separation} phenomenon relies crucially on the ability of social networks to mobilize in a targeted manner, using geographical information in recruiting participants. The data provides significant support for the presence of geographical targeting, even under time pressure. In fact, we observe that targeting increases as a function of time pressure, as the challenge approaches its deadline. 

We were also able to quantify the intensity of targeted mobilization behavior, in comparison with the baseline of untargeted flow of global social media communication. This supports the general notion that social networks are able to tune their geographical communication to suit the task at hand. For example, using Twitter data, it was shown that the Occupy Wall Street social movement in the United States exhibits significant localization (at the state level) when it comes to messages that facilitate resource mobilization and coordination, with reference protest action and specific places and times. In contrast, information flows across state boundaries are more likely to contain framing language to develop narrative frames that reinforce collective purpose at the national level \cite{conover2013geospatial,conover2013digital}. Our findings complement these results, by contributing towards a general theory that link the purpose of social mobilization to the temporal and spatial dynamics of different forms of communication.

Within high volume social media communications, considerable effort is required to persuade people about the importance of a particular message or cause or even to notice it at all. Both considerations are crucial for a successful mobilisation process. Previous work has shown that shared news stories of interest become obsolete on a timescale $\approx\mathrm{1h}$~\cite{WuNovelty} and that the amount of cognitive resources an individual dedicates to online communications is limited and inelastic~\cite{miritello2013}, meaning that the intrinsic importance of the message cannot be relied upon to overcome informational overload and to motivate its sharing. In addition, active interaction with a task requires much more attentional cost to an individual than simple observation~\cite{centreOfAttention} and connected individuals vital for propagation also have an associated high inertia~\cite{HodasVisibility}. The importance of targeted personal interactions (typified by Twitter \texttt{@} mentions) can be seen in this context; personalised messages obligate greater cognitive effort from the receiver overcoming the inevitable slide into obsolescence of a single subject over time. Geographical targeting now has an additional advantage beyond the increased chance of recruiting a first hand searcher as the targeting converges; increased personal affiliation of the receiver with the message. The empirical evidence presented above suggests that large distributed communities intuitively understand these considerations and can leverage them in a timely and powerful manner.

\section*{Materials and Methods}

\section{Twitter} The Web site Twitter is an extremely popular micro-blogging service which also incorporates a social network. Users create short messages ('\emph{tweets}') of 140 characters or less which contain text and/or shortened hyperlinks to other webpages or images of interest. Users tweets appear in the \emph{feed} of all other users who have chosen to \emph{follow} her. A user may also opt to make the content of their tweets visible to the public. \emph{Tweets} contain \emph{hashtags} to signify that the tweet is relevant to a particular topic i.e. \texttt{\#playTag} was a popular hashtag for the Tag Challenge. Users may also choose to target a Tweet to a particular user, regardless of whether the users are connected by a follower/following link, rather than simply broadcasting to her followers. This is done by including a user's \emph{Twitter handle} e.g. \texttt{@crowdscannerhq}.

We collected the full set of relevant tweets from the period 13$^{\textrm{th}}$ February to 10$^{\textrm{th}}$ April using a paid service~\cite{hootsuite} according to appropriate hash tags and keywords or targeted mentions (\emph{@} mentions) of competing teams. Tweets originating from \texttt{@TagTeam\_}, \texttt{@CrowdscannerHQ}, \texttt{@TagChallenge}, \texttt{@Tagteam} and \texttt{@Tag\_Challenge} were discarded. Tweets from the participating teams were excluded from these daily totals since the teams had an interest in increasing the daily tweet volumes. The tweets were then manually filtered for relevance by relevant hashtags such as \texttt{\#playTag}, \texttt{\#tagchallenge}, \texttt{\#tag} and any links contained within the tweet. 1263 tweets out of 2181 remained after the filtering process. 

Tweets from users with no reliable location information which could be geo-coded were discarded, further care was taken to recognise and eliminate artefacts of the geocoding process which led to spurious latitude/longitude coordinates. e.g. `The world' becoming `(0.0,0.0)'. Tweets originating from within 25km of the defined city centres~\cite{geohack} were considered to originate from the city.

\section{Facebook} As the largest Web-based social network in the world, Facebook has over 1 billion active users. The daily number of impressions were sourced using the Facebook Insights Application Programming Interface (API)~\cite{facebook}. This covers any user engagement with Tag Challenge page, such as posts on one's ``wall'' or expressions of approval by friends using the ``like'' button, etc.

\section{Google Analytics} The traffic to the official website was recorded between 14$^{\mathrm{th}}$ February and 4$^{\mathrm{th}}$ April. A total of 1000 unique users and their IP addresses were recorded in this period. We used an online service~\cite{ipinfo} to derive approximate location coordinates from this IP. To mitigate the effect of noise due to the variable volumes of traffic, a moving average was taken for each day, using a sliding window defined as $\textstyle(\textrm{MA}(\textrm{prop}^{\beta}(t))_{n}=(\textrm{prop}^{\beta}(t-n)+...+\textrm{prop}^{\beta}(t-1)+\textrm{prop}^{\beta}(t))/n$, where $\textrm{prop}^{\beta}(i)$ is the proportion of distance ordered tweets within the $\beta$-th percentile on day $i$ which were within a tag city and $n$ is the order of the moving average. Figure 2 corresponds to $n=4$ and $\beta=0.25$.

Even the full set of unsmoothed data ($n=0$,\,$\beta=1$) reveals a geographically convergent trend $(r,p)=(-0.34, <10^{-5})$). We excluded tweets from the Tag teams since the teams may have actively pursued a strategy of geographical convergence skewing the results.

\section{Simulation} A coarsened network of air travel connections was constructed as follows. Firstly the largest 280 Metropolitan Statistical Areas (MSA) were considered across all continents. Polycentric MSAs such as the New York Tri-state area in New York were collapsed into one node in the network. A full list of global airports and connections between them was taken from Open Flights~\cite{openflights}. In order to coarsen the data, airports were agglomerated to the geographically closest MSA using open data~\cite{geonames}~\cite{geographytwitter}. Now the many airports of Greater London; Heathrow, Stanstead, Luton, Gatwick etc are all considered together. This coarsening helps mitigate the effect of anomalous behavior within sparsely populated regional clusters with unusual locality, such as Alaska~\cite{amaralAirports}.

The network edge weights are based on a normalised number of flights between every 2 cities, with self loop weights set to 0.39 representing the probability of communication within the same MSA \cite{geographytwitter}. We construct an adjacency matrix representation of the network, namely an $n \times n$ square matrix $A$, where $n$ is the number of MSAs, and $A_{ij}$ is the weight of the directed edge between cities $i$ and $j$. The adjacency matrix was row normalised, such that row $A_i$ represents a probability distribution over the target node reached by a random walker leaving node $i$. This results in an adjacency matrix which is nearly symmetric.

We then simulated a random walk over this network. With probability $g \in [0,1]$, so called greediness bias, we move towards the closest Tag Challenge cities. And with probability $1-g$ we take a pure random walk with probabilities proportional to the outgoing edge weights. A random walk, with $g=0$ corresponds to the eigenvector centrality vector of the different MSAs (see SI Appendix for further details).

\section*{Acknowledgments}

We would like to thanks David Alan Grier for discussion. We are grateful to all our volunteers that joined the CrowdScanner team.

\bibliography{scibib}

\section*{Figure Legends}

\begin{figure}[ht]
\begin{center}
\includegraphics[width=0.9\textwidth]{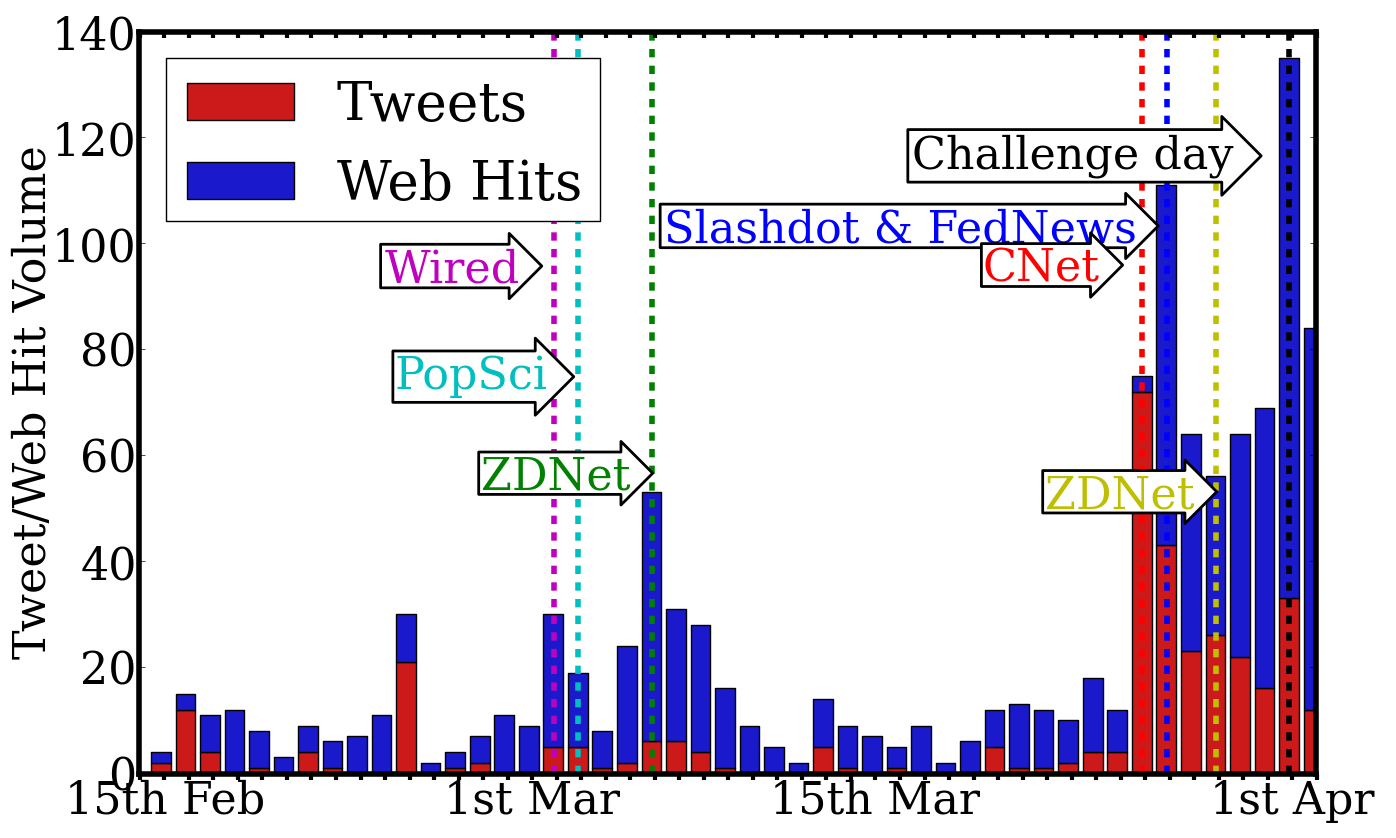}
\caption{Daily volumes of Tag Challenge related Tweets and Web
  hits on \texttt{http://www.tag-challenge.com} up to the challenge day. Major media coverage events are highlighted.}
\end{center}
\label{fig:volume}
\end{figure}

\begin{figure}[ht]
\begin{center}
\includegraphics[width=0.9\textwidth]{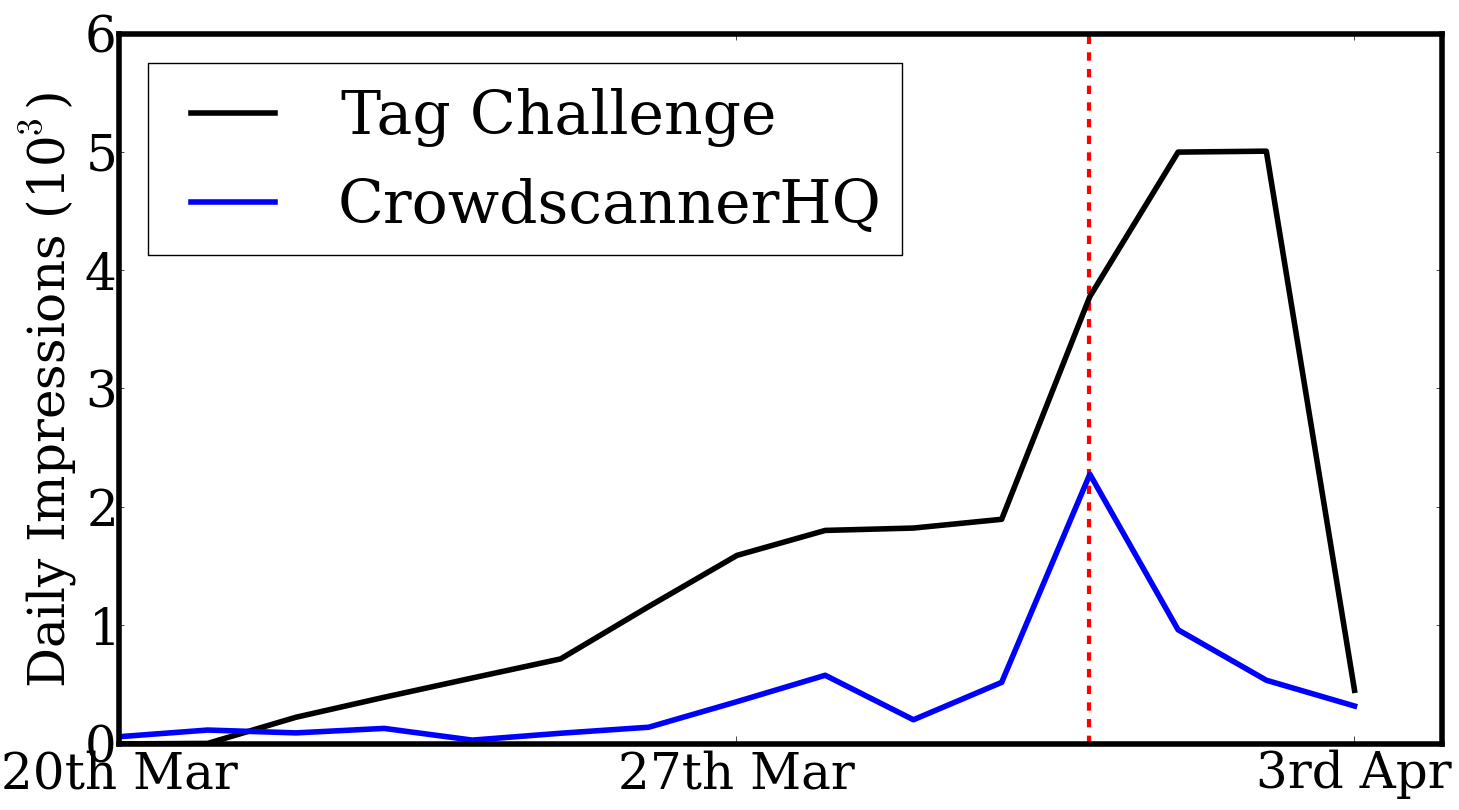}
\caption{Daily number of impressions on Facebook for the winning team \texttt{CrowdscannerHQ}, and the official Tag Challenge organizers. The vertical dotted line denotes the release of the first mug shots.}
\end{center}
\label{fig:impressions}
\end{figure}

\begin{figure}[ht]
\begin{center}
\includegraphics[width=0.9\textwidth]{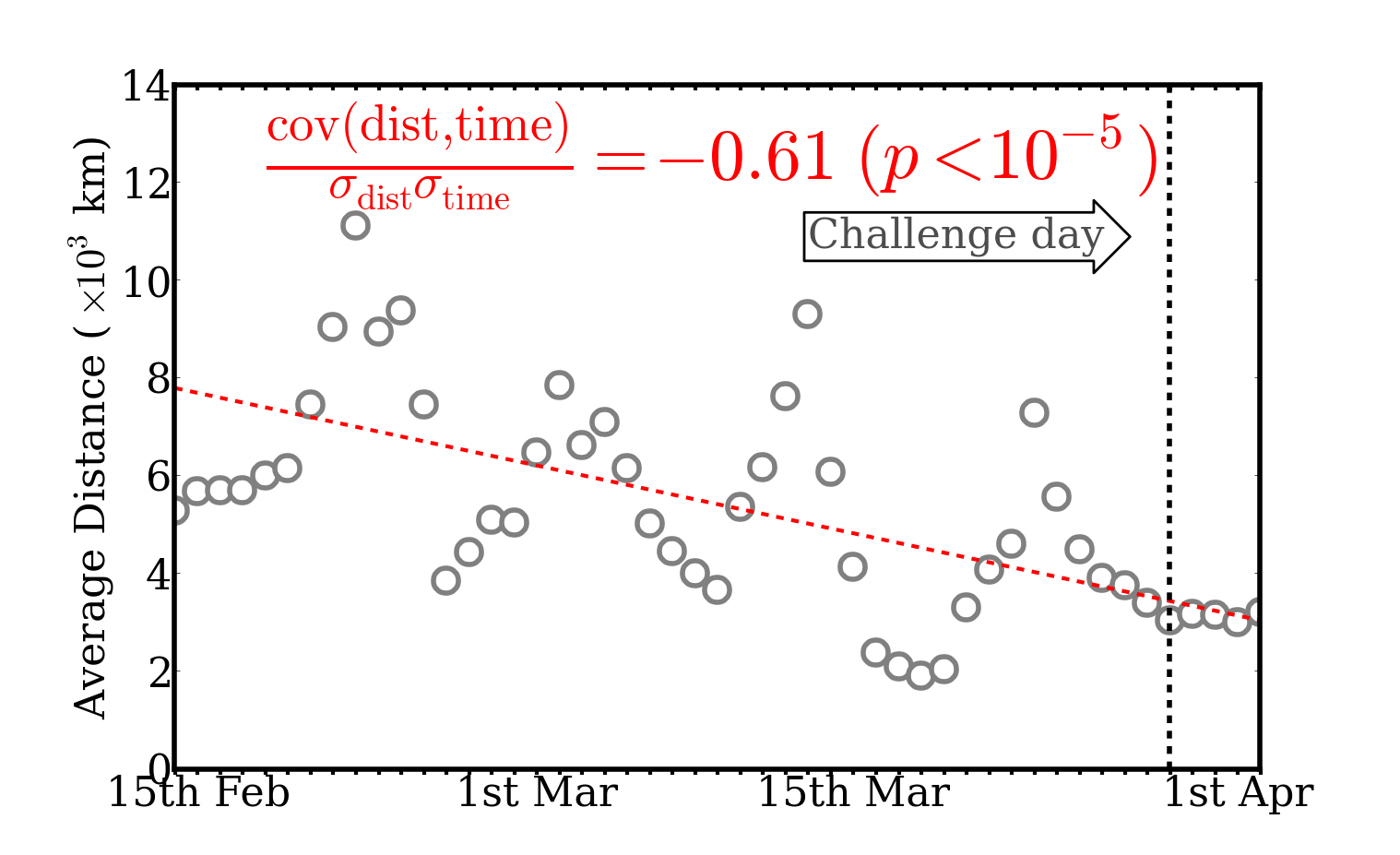}
\caption{Distance convergence toward Tag Challenge cities of web hits on
  \texttt{http://www.tag-challenge.com}. We consider a moving average of distance filtered daily tweet traffic (MA(prop$^{0.25}$(t))$_{4}$) (grey circles), which is fit with a linear regression (red line) giving a correlation of $(r,p)=(-0.61, <10^{-5})$.}
\end{center}
  \label{fig:convergence}
\end{figure}

\begin{figure}[ht]
\begin{center}
\includegraphics[width=0.9\textwidth]{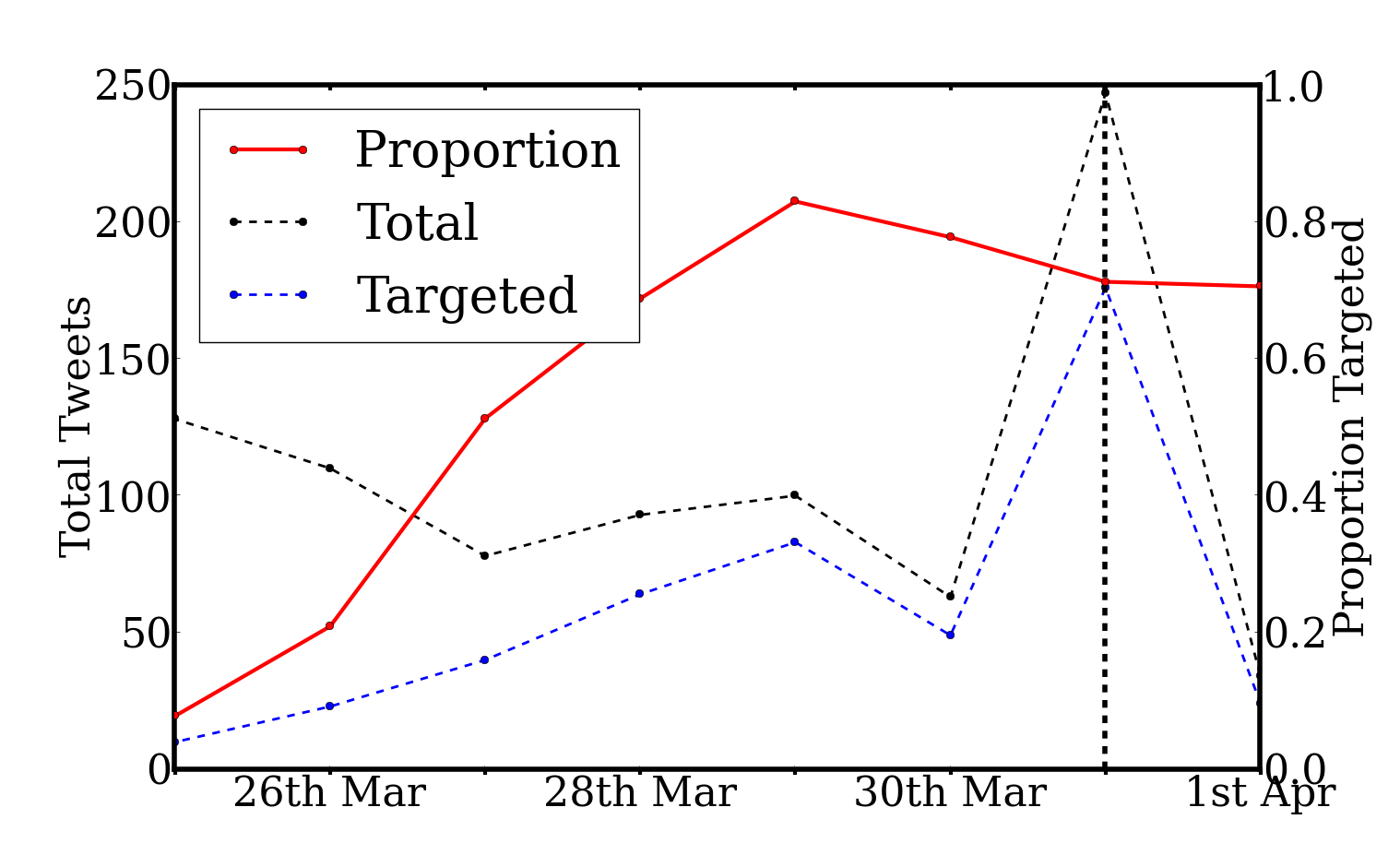}
\caption{The total daily number of Tweets (black line), the number targeting individuals via @-mentions (blue line) and their proportion (red line). Correlation of targeted proportion with time was found as $(r,p)=(0.825,0.012)$
}
\label{fig:propotion}
\end{center}
\end{figure}

\begin{figure}[ht]
\begin{center}
\includegraphics[width=0.9\textwidth]{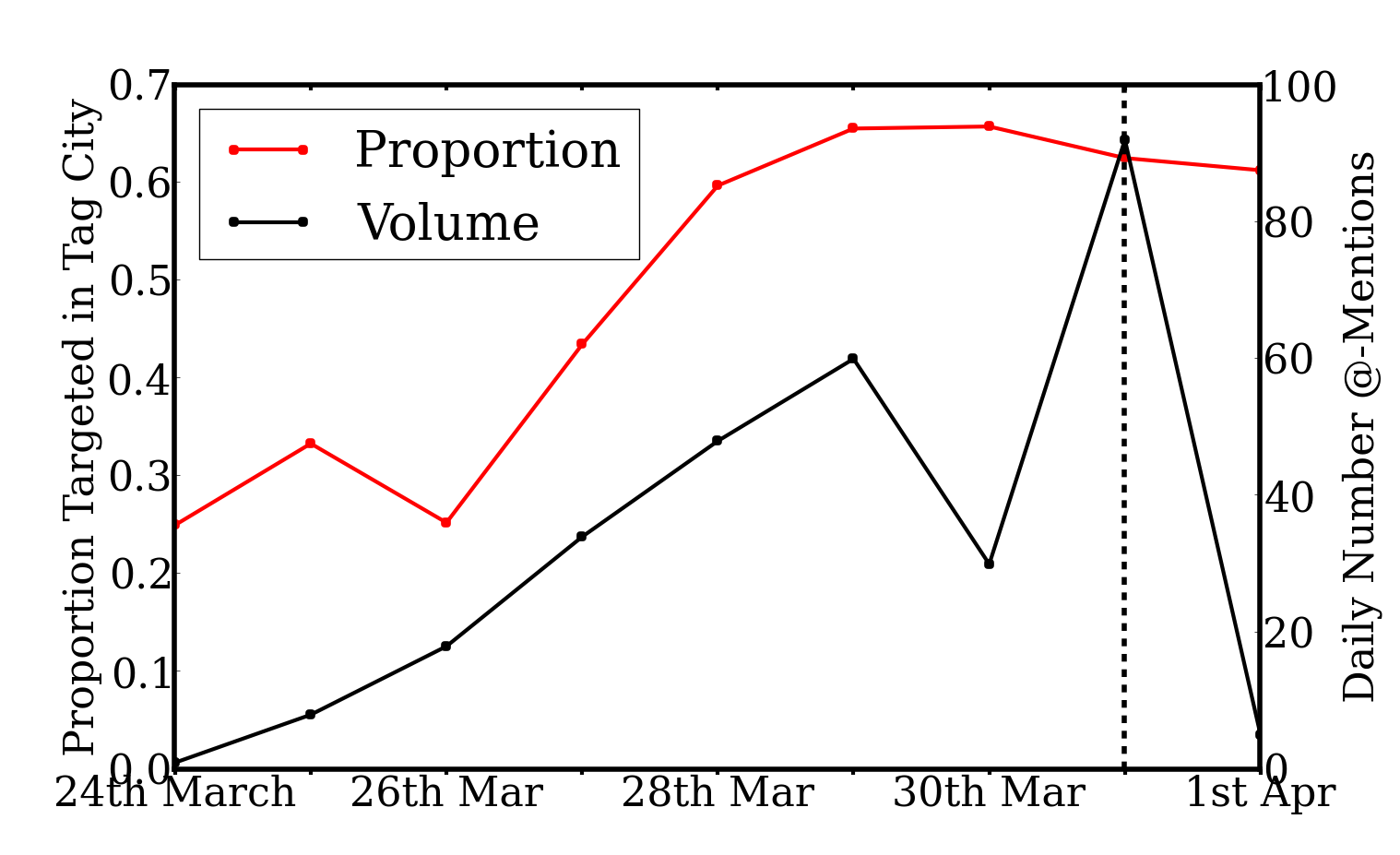}
\caption{Daily proportion of \texttt{@}-mentioned users which are located within a tag city. Noise is eliminated by smoothing with a 4 day moving average. Correlation with time reveals a trend given by $(r,p)=(0.912,0.002)$}
\end{center}
\label{fig:mentioned}
\end{figure}

\begin{figure*}[ht]
\begin{center}
\includegraphics[width=0.95\textwidth]{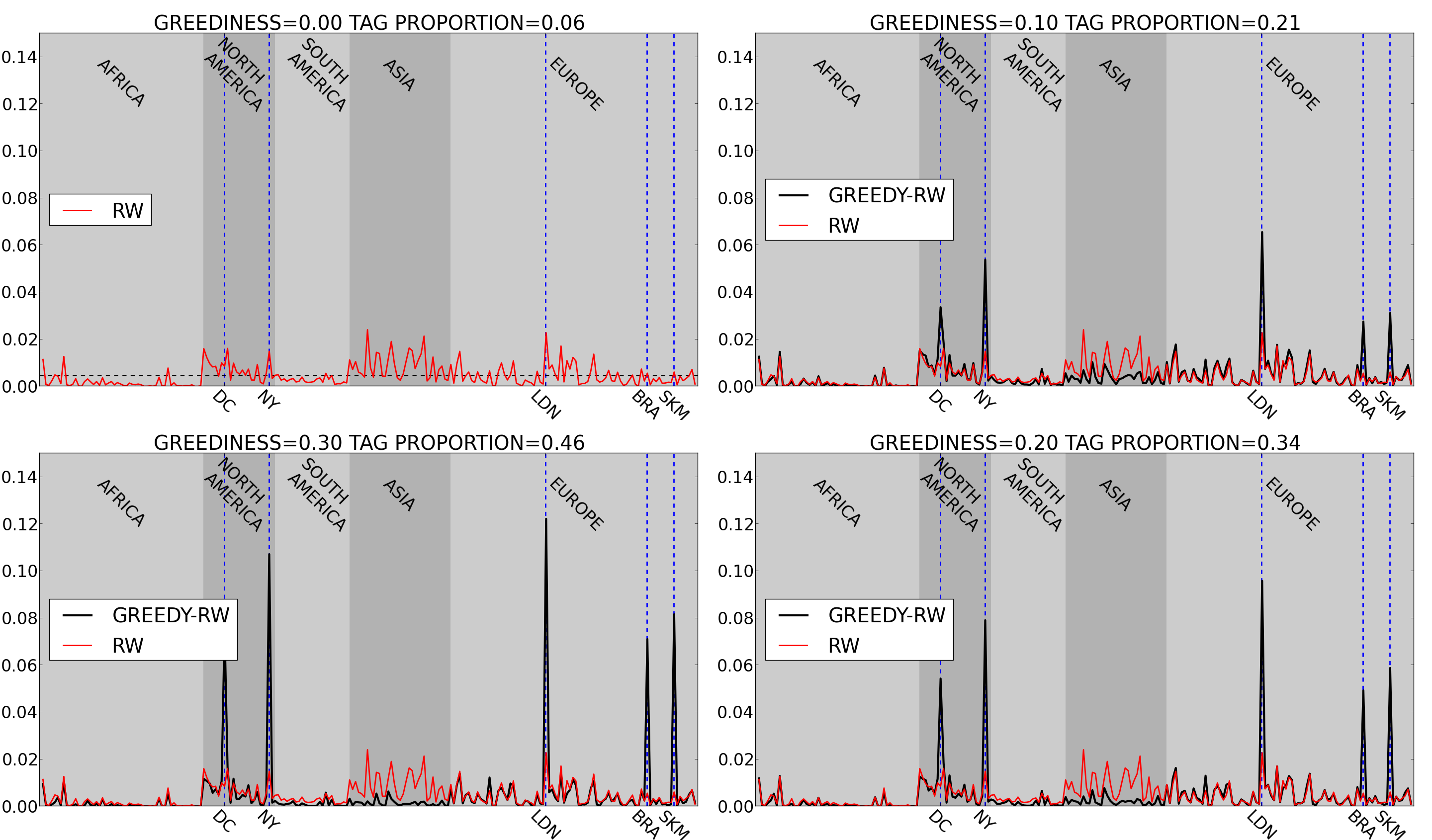}
\caption{Plot of stationary distribution during a random walk on global MSA network, with increasing degree of greediness (targeting) moving clockwise from top left. The red line represents an pure, untargeted random walk, corresponding to pure random mobilization via broadcast messaging. (Top left) The horizontal dashed line represents the uniform distribution of centralities expected in a fully connected graph. The black line in other plots represents a greedy random walk. (Bottom right) When the greediness is increased to $30\%$ we match the observed proportion of targeted messages reaching the Tag Challenge cities. The shading represents MSAs from different continents. The 5 tag cities are marked with vertical, dashed blue lines.}
\label{fig:tile_full}
\end{center}
\end{figure*}

\begin{figure}[ht]
\begin{center}
\includegraphics[width=0.9\textwidth]{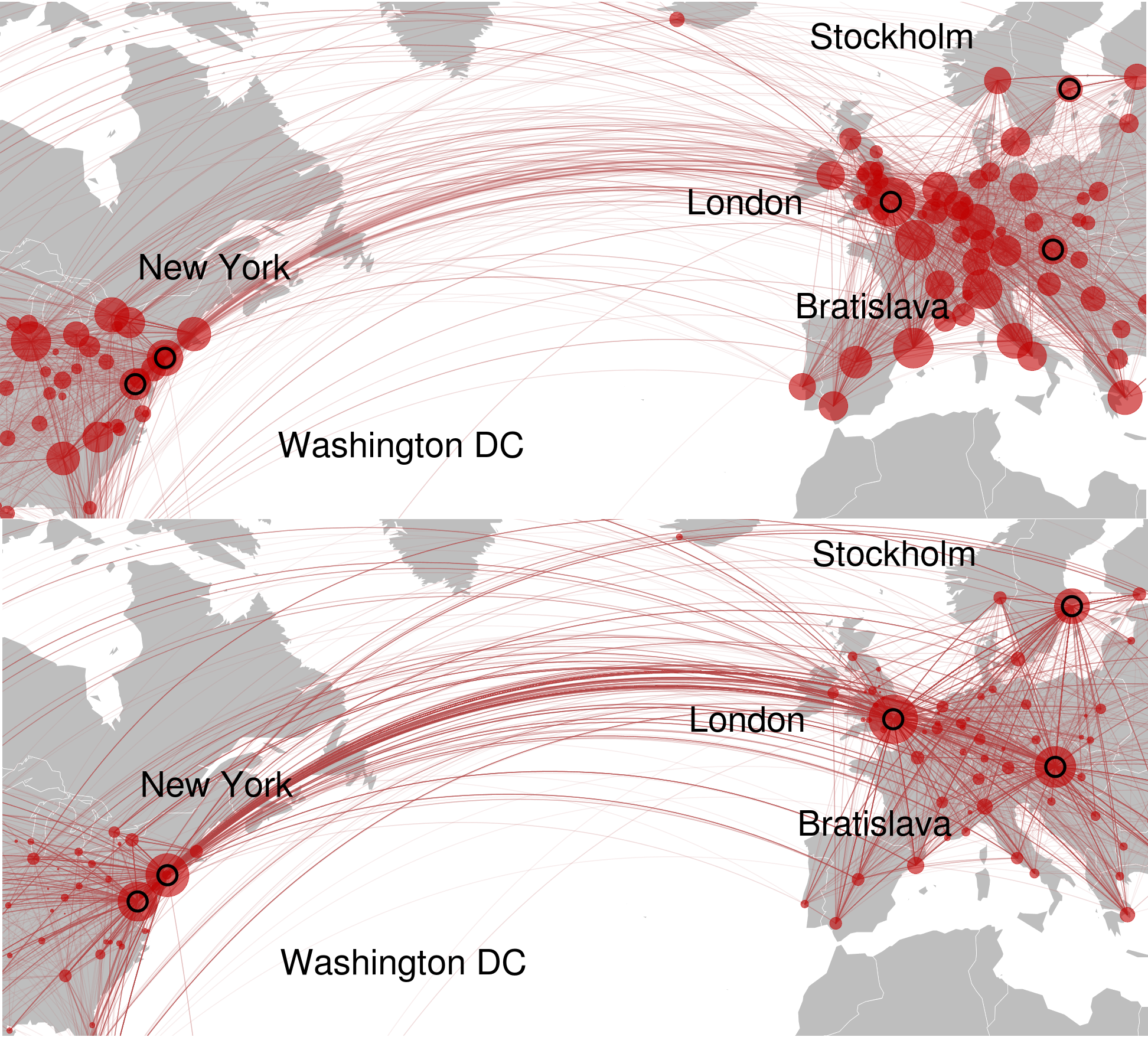}
\caption{Map showing communication network within Europe and North America following an unbiased random walk (upper) and under 30\% targeting (lower). The area of red circles are proportional to centrality.}
\end{center}
\label{fig:centralityMap}
\end{figure}

\end{document}


\noindent {\LARGE Supporting online material for:}\\[0.5cm]
\noindent{\Large Targeted Social Mobilization in a Global Manhunt}
\appendix

\tableofcontents
\setcounter{tocdepth}{2}

\section{Flight Network}

\begin{figure}[ht!]
\begin{center}
\includegraphics[width=\textwidth]{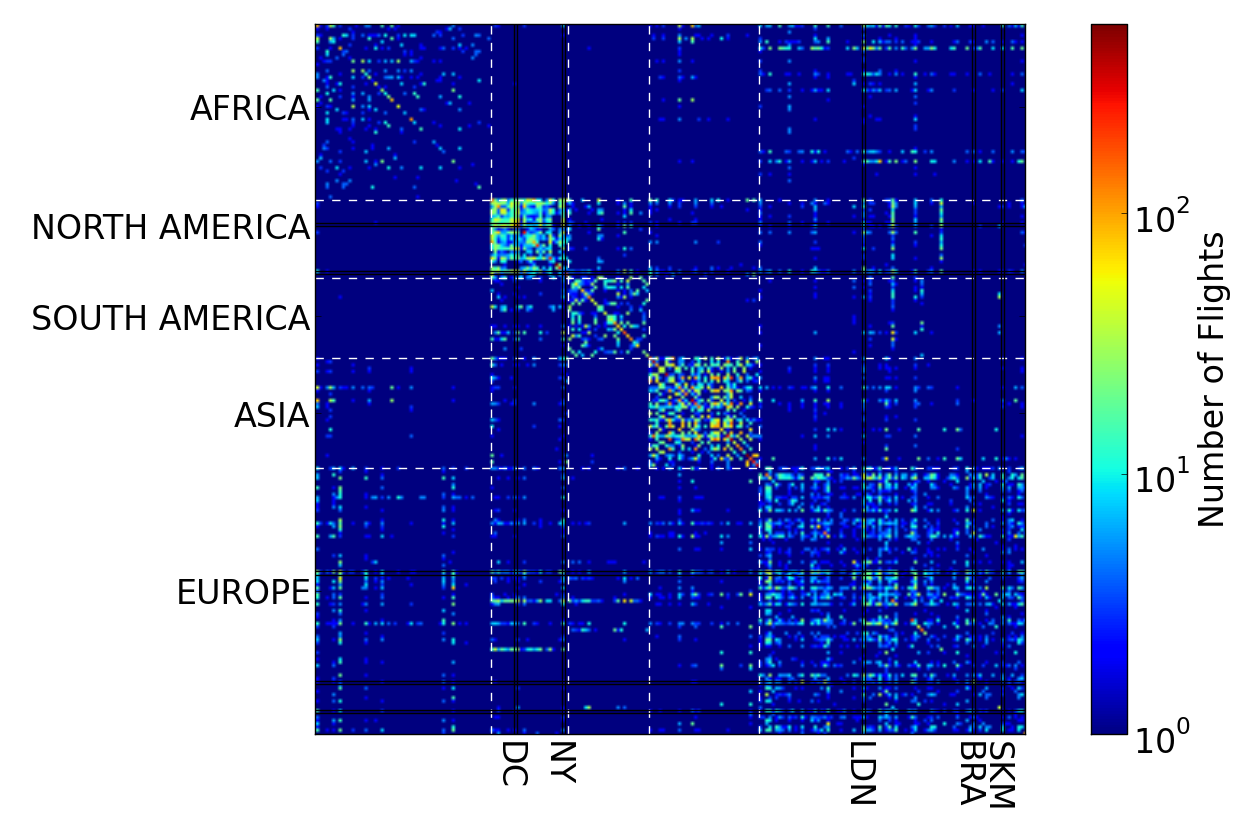}
\caption{Heat map of raw flight numbers. Continent limits are marked by white dashed lines and tag cities with black lines.}
\label{raw}
\end{center}
\end{figure}

\begin{figure}[ht]
\begin{center}
\includegraphics[width=\textwidth]{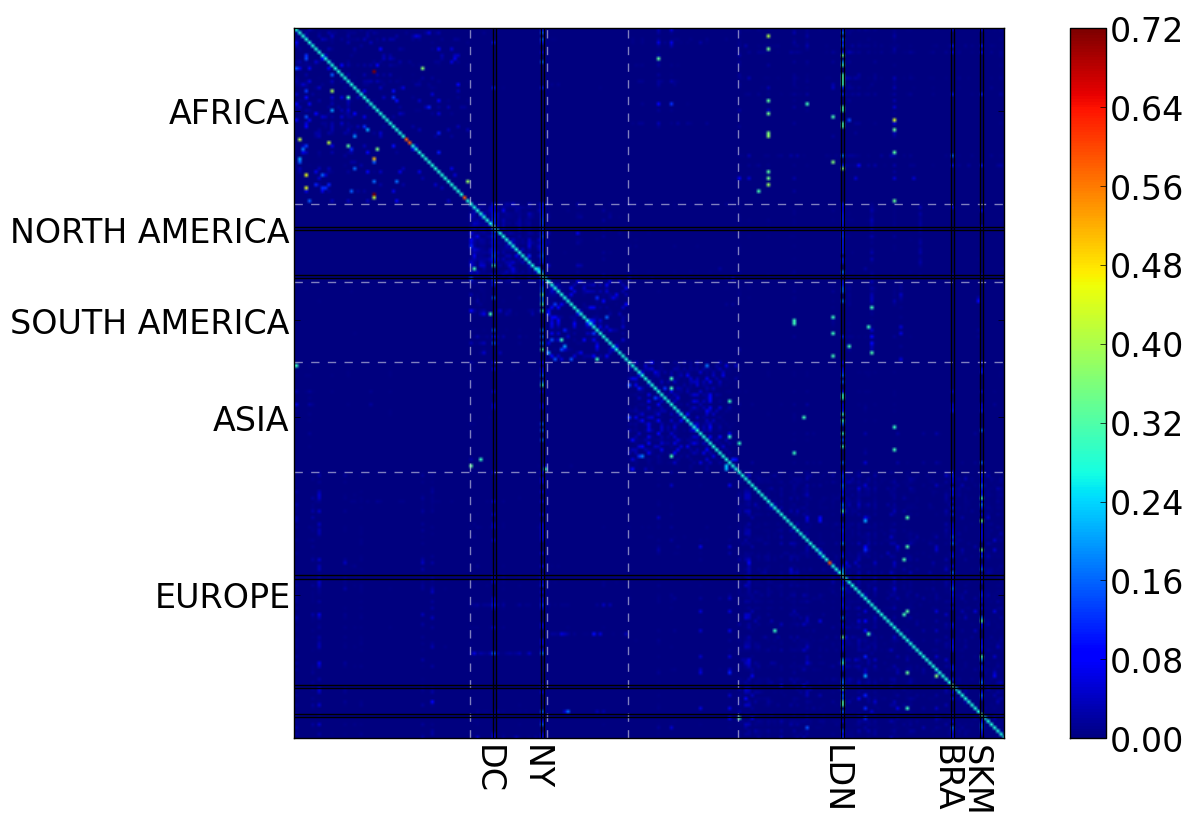}
\caption{Heat map of normalised and locality-adjusted adjacency matrix with greediness set to 0.3. Continent limits are marked by white dashed lines and tag cities with black lines.}
\label{greedy}
\end{center}
\end{figure}

We can visualise the adjacency matrix of the MSA network both in terms of raw number of flights (Fig (\ref{raw})) and the normalised, locality and greediness-adjusted edges (Fig (\ref{greedy})). There are 2 striking features in Figure (\ref{raw}), firstly we see a strong community structure with respect to continents as also observed in~\cite{amaralAirports}, particularly within Asia and secondly the high occupation of the diagonal. While the intercontinent connectivity is intuitively understandable, the latter is explained by geography. In regions such as Polynesia, there are a large number of flights between small regional airports on different islands but few outside of the community. Among the MSA's which represented the largest number of airports, were Jakarta (Indonesia), Auckland (New Zealand), Anchorage (Alaska, USA) and Port Moresby (Papua New Guinea) which are all regional hubs within sparsely-populated or archipelagic areas which may only feasibly be navigated by air. Since these small regional airports all agglomerate to a single MSA, it appears that a large volume of flights appear to leave from and arrive at the same MSA. Therefore these hubs have large unadjusted localities represented by large values along the diagonal. This artifact of the agglomeration process has a negligble affect on the structure of the network as a whole since these communities are not particularly central; this can be seen by the low centralities of these MSA's\\

\begin{table}[ht!]
\caption{Table of regional hub MSA's and centralities}
\begin{center}
\begin{tabular}{|c|c|c|}
	\hline
	MSA & Centrality & Centrality/Centrality$_{\mathrm{equal}}$\\
	\hline
	Jakarta & 0.00572 & 1.24\\
       Auckland & 0.00289 & 0.63\\
	     Anchorage & 0.00092 & 0.20\\
	  Port Moresby & 0.00176 & 0.38\\
	\hline
	Equal & 0.0046 & \\
	\hline

\end{tabular}
\end{center}
\end{table}

The adjusted adjacency matrix used in the simulations and shown in Figure (\ref{greedy}) maintains the strong continental community effect, however the localities have been uniformly set to 0.39 and a greediness of 30\% has been applied. In a few cases this greediness leads to increased locality if all outgoing edges from an MSA move the message away from the nearest Tag city. The greediness also leads to a number of strong connections to Europe (but not directly the the European Tag cities); the turquoise dots representing strength 0.3 in the columns on the right of the figure.

Table (\ref{centralityTable}) shows the centralities of the most central MSAs in the network along with the Tag cities for comparison. All of the tag cities are above the baseline of equal centrality amongst all the nodes, however London, Washington DC and NY are expecially so.

\begin{table}
\caption{Table of MSA's with highest centrality values after locality adjustment (and tag cities for comparison)}
\begin{center}
\begin{tabular}{|c|c|c|}
	\hline
	MSA & Centrality & Centrality/Centrality$_{\mathrm{equal}}$\\
	\hline
	Shanghai & 0.02242 & 6.28\\
	  London & 0.02231 & 6.25\\
      Chongquing & 0.01988 & 5.57\\
		Beijing & 0.01745 & 4.89\\
	      Barcelona & 0.0155 & 4.34\\
	 Milan & 0.0149 & 4.17\\
	
	\hline
	London & 0.02225 & 6.25\\
	    NY & 0.010 & 2.80\\
	    DC & 0.00645 & 1.87\\
	Bratislava/Vienna & 0.0054 & 1.51\\
		Stockholm & 0.0059 & 1.65\\
	\hline
	Equal & 0.00357 & \\
	\hline
\end{tabular}
\label{centralityTable}
\end{center}
\end{table}
\clearpage

\section{Reduced Network}

\begin{figure}[ht]
\begin{center}
\includegraphics[width=0.95\textwidth]{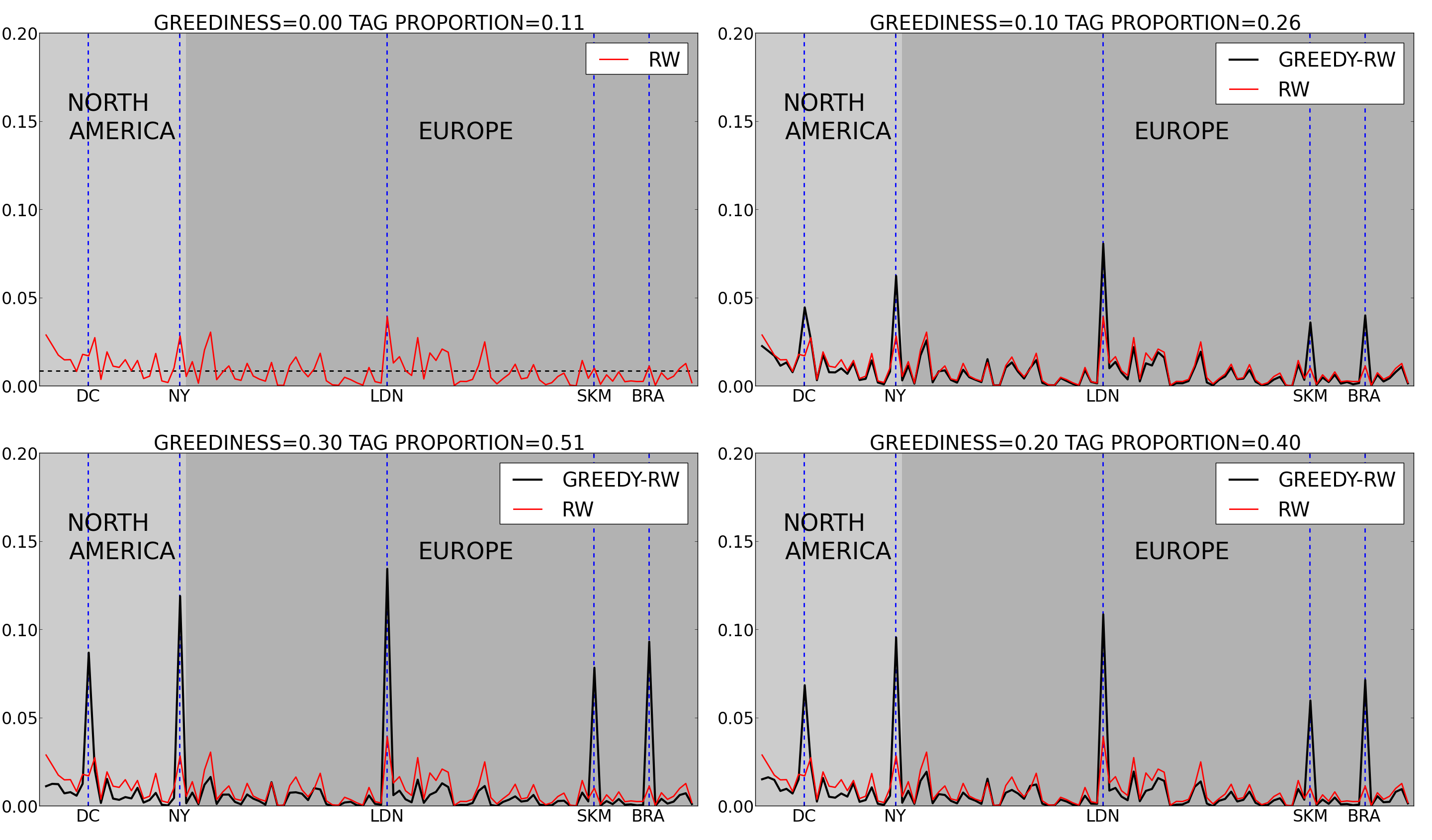}
\caption{Plot of stationary distribution during a random walk on reduced MSA network (comprising only North America and Europe), with increasing degree of greediness moving clockwise from top left. The red line represents an unbiased random walk, corresponding to pure random mobilization via broadcast messaging. (Top left) The horizontal dashed line represents the uniform distribution of centralities expected in a fully connected graph. The black line in other plots represents a greedy random walk.}
\label{fig:tile_full}
\end{center}
\end{figure}

Figure (\ref{fig:tile_full}) shows the effective centralities of the cities within a reduced network comprising the cities of North America and Europe only (compare with the full global network shown in Fig (6) in the main paper). A degree of targeting of 30\% now leads to a proportion of messages reaching the tag cities of 0.50 (compared to 0.44 using the full network). As expected the proportion of time spent in the tag cities increases as nodes are removed from the network. In fact the effect of the removal of the South American, African and Asian MSA's becomes smaller as targeting becomes stronger and routes the message towards the western hemisphere. Considering the pure, non-targeting random walk the reduced network increases the Tag proportion from 0.05 to 0.09; a percentage increase of 80\%. However as the targeting becomes stronger this percentage difference becomes smaller. When greediness is set to 30\% the reduced network increases the tag proportion from 0.44 to 0.50, an increase of only 13\%.

\clearpage

\section{Website Traffic}

\begin{figure}[ht]
\begin{center}
\includegraphics[width=.9\textwidth]{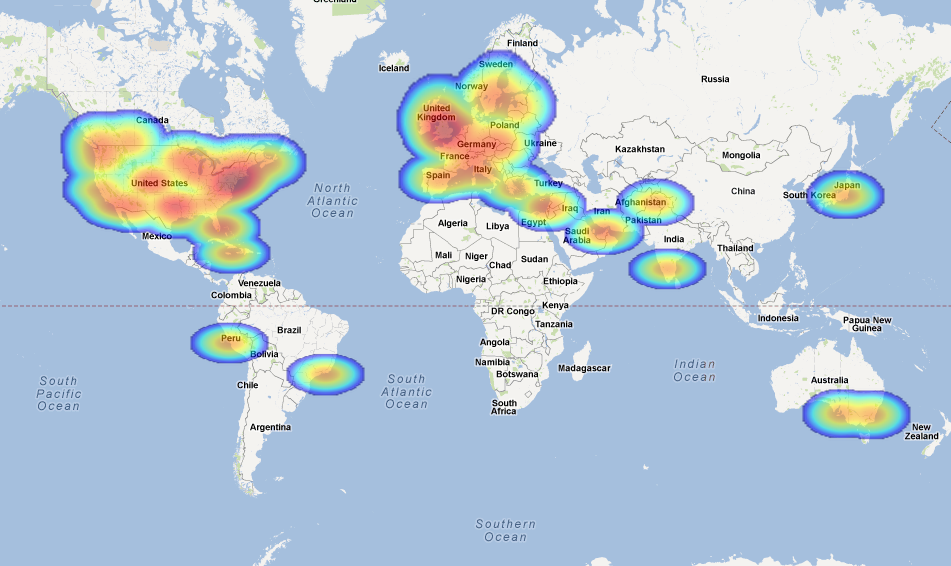}
\end{center}
\caption{Heatmap showing traffic to crowdscanner.com on 48 hours approaching the challenge.}
\label{statcounter_heatmap}
\end{figure}

Figure (\ref{statcounter_heatmap}) shows the geographical distribution of traffic to our team's website in the 48 hours approaching the challenge. Traffic overwhelmingly originates from Europe and North America, particularly since this snapshot is from the critical latter stages of the propagation process, but we can also notice the presence of traffic originating from South America, Australia and Asia Pacific. The fact that tag traffic is significant even outside the Anglosphere suggests that the information diffusion either took place in languages other than English (a small but significant number of tweets were in languages other than English) or the English language media exposure was accessible via the \emph{lingua franca}. While the South American, Asian and African nodes clearly participated in the diffusion, the network upon which this took place is likely somewhere between the reduced network presented here and the full global network presented in the main paper. Regardless of which extreme of network substrate dominates, we can conclude that significant targeting is required to reproduce the proportions of traffic towards the Tag cities.

\section{List of Cities}

\begin{table}
\caption{List of cities in North America}
\begin{center}
\begin{tabular}{|c|c|}
	\hline
 Los Angeles&Chicago\\
   Toronto&Dallas\\
     Houston&Philadelphia\\
      Detroit&Washington\\
        Atlanta&San Diego\\
	 Boston&Riverside\\
	   Phoenix&Seattle\\
	     Minneapolis&Tampa\\
	       Saint Louis&Baltimore\\
		 Denver&El Paso\\
		 Anchorage&Vancouver\\
		  New York&Honolulu\\
 San Antonio&San Jose\\
Jacksonville&Indianapolis\\
      Austin&San Francisco\\
    Columbus&Fort Worth\\
   Charlotte&Detroit\\
     El Paso&Memphis\\
      Boston&Nashville\\
      Denver&Baltimore\\
  Louisville&Milwaukee\\
    Portland&Oklahoma City\\
   Las Vegas&Albuquerque\\
      Tucson&Fresno\\
  Sacramento&Long Beach\\
 Kansas City&Virginia Beach\\
     Corpus Christi&Colorado Springs\\
     Raleigh&Omaha\\
       Miami&Tulsa\\
     Oakland&Cleveland\\
 Minneapolis&Wichita\\
 New Orleans&Bakersfield\\
       Tampa&Honolulu\\
   Santa Ana&St. Louis\\
  Pittsburgh&Lexington\\
  Cincinnati&Anchorage\\
  Saint Paul&Greensboro\\
     Lincoln&Buffalo\\
  Fort Wayne&Saint Petersburg\\
     Orlando&Norfolk\\
      Laredo&Madison\\
     Lubbock&Durham\\
WinstonSalem&Garland\\
 Baton Rouge&Reno\\
  Scottsdale&San Bernardino\\
  Birmingham&Rochester\\
	\hline
\end{tabular}
\end{center}
\end{table}
\begin{table}
\caption{List of cities in South America}
\begin{center}
\begin{tabular}{|c|c|}
	\hline
	City&Country\\
	\hline

	Mexico City&Mexico\\
  S\^{a}o Paulo&Brazil\\
Buenos Aires&Argentina\\
   Rio de Janeiro&Brazil\\
	     Lima&Peru\\
 Bogot\'{a}&Colombia\\
   Caracas&Venezuala\\
  Santiago&Chile\\
	Belo Horizonte&Brazil\\
	   Guadalajara&Mexico\\
	     Monterrey&Mexico\\
	Porto Alegre&Brazil\\
	      Recife&Brazil\\
	    Salvador&Brazil\\
	   Brasilia&Brazil\\
	   Medellín&Colombia\\
	   Fortaleza&Brazil\\
	      Santo Domingo&Dominic Republic\\
			  Curitiba&Brazil\\
	     Guatemala City&Guatemala\\
			      Cali&Colombia\\
		  Guayaquil&Ecuador\\
		     Puebla&Mexico\\
		   Campinas&Brazil\\
		   San Juan&Puerto Rico\\
	
	\hline
\end{tabular}
\end{center}
\end{table}
\begin{table}
\caption{List of cities in Europe}
\begin{center}
\begin{tabular}{|c|c|c|c|}
	\hline
	City&Country&City&Country\\
	\hline
	Amsterdam  & Netherlands&Antwerp & Belgium\\
     Athens & Greece&  Barcelona &  Spain\\
       Baku & Azerbaijan&   Belgrade & Serbia\\
     Berlin & Germany&     Bremen & Germany\\
    Bristol & United Kingdom&   Brussels &Belgium\\
  Bucharest &  Romania&  Budapest &  Hungary\\
    Cardiff & United Kingdom& Copenhagen &  Denmark\\
    Donetsk & Ukraine&Dnipropetrovsk & Ukraine\\
    Dublin  & Ireland&     Frankfurt & Germany\\
    Glasgow & United Kingdom& Hamburg & Germany\\
  Helsinki  & Finland&      Istanbul & Turkey\\
  Katowice  & Poland&	 Kazan & Russia\\
    Kharkiv & Ukraine& Kiev &  Ukraine\\
     Krakow &  Poland&  Leeds & United Kingdom\\
    Lisbon  & Portugal&Liverpool & United Kingdom\\
\L\'{o}d\'{z} & Poland&       London & United Kingdom\\
		Lyon & France&Madrid &  Spain\\
Manchester & United Kingdom& Marseille & France\\
    Milan  & Italy&     Minsk & Belarus\\
    Moscow &  Russia&    Munich & Germany\\
   Napoli  & Italy&      Nice & France\\
Nizhniy Novgorod & Russia&      Nottingham & United Kingdom\\
       Nuremberg & Germany&	  Odessa & Ukraine\\
	    Oslo & Norway&   Paris  & France\\
     Perm & Russia&    Porto & Portugal\\
   Portsmouth & United Kingdom&       Prague & Czech Republic\\
D\"{u}sseldorf & Germany&       Cologne & Germany\\
			Riga & Latvia&	 Roma  & Italy\\
 Rostov-on-Don & Russia&	    Rotterdam & Netherlands\\
	     Dortmund & Germany&	 Saarbr\"{u}cken & Germany\\
     Saint Petersburg & Russia&		      Samara & Russia\\
		     Saratov & Russia&	      Seville & Spain\\
		       Sofia & Bulgaria&     Stockholm & Sweden\\
	    Sheffield &United Kingdom&	    Stuttgart & Germany\\
	      Tbilisi & Georgia&	 Thessaloniki &  Greece\\
		       Turin & Italy&     Newcastle & United Kingdom\\
			 Ufa & Russia&      Valencia & Spain\\
	Vienna & Austria&	    Volgograd & Russia\\
	Warsaw &  Poland&	    Birmingham& United Kingdom\\
       Yerevan & Armenia&	       Zagreb & Croatia\\
	      Zurich  & Switzerland&	   Reykjavík &Iceland\\

		       Milan & Italy&Katowice & Poland\\
		       Lille & France& &\\
	\hline
\end{tabular}
\end{center}
\end{table}
\begin{table}
\caption{List of cities in Asia}
\begin{center}
\begin{tabular}{|c|c|c|c|}	
	\hline
	City&Country&City&Country\\
	\hline
	Tokyo& Japan&Jakarta &Indonesia\\
 Seoul& South Korea&Delhi& India\\
Mumbai &India&Manila &Philippines\\
Shanghai &China&Osaka &Japan\\
 Kolkata &India&Karachi &Pakistan\\
Guangzhou& China&Dhaka &Bangladesh\\
 Shenzhen& China&Tehran& Iran\\
  Beijing& China&Bangkok& Thailand\\
  Chennai& India&Nagoya &Japan\\
Bangalore &India&Hong Kong&Hong Kong\\
    Wuhan &China&Taipei &Taiwan\\
   Lahore &Pakistan&Tianjin& China\\
Kuala Lumpur& Malysia&Chongqing &China\\
  Hyderabad &India&Ho Chi Min City&Vietnam\\
    Shenyang& China&Ahmedabad &India\\
   Brisbane &Australia&Samarkand &Uzbekistan\\
   Auckland &New Zealan&Port Moresby&Papua New Guineau\\
     Sydney &Australia&Melbourne &Australia\\
      Perth &Australia&Adelaide &Australia\\
	\hline
\end{tabular}
\end{center}
\end{table}
\begin{table}
\caption{List of cities in Africa}
\begin{center}
\begin{tabular}{|c|c|}
	\hline
	City&Country\\
	\hline
	Cairo & Egypt\\
	Lagos & Nigeria\\
	Kinshasa & Democratic Republic of the Congo\\
    Johannesburg & South Africa\\
	       Khartoum & Sudan\\
	     Alexandria & Egypt\\
		Abidjan & Ivory Coast\\
      Casablanca & Morocco\\
	      Cape Town & South Africa\\
		 Durban & South Africa\\
	   Accra & Ghana\\
	 Nairobi & Kenya\\
	  Ibadan & Nigeria\\
   Dar es Salaam & Tanzania\\
		Algiers & Algeria\\
     Addis Ababa & Ethiopia\\
		 Luanda & Angola\\
	   Dakar & Senegal\\
	Pretoria & South Africa\\
	 Tripoli & Libya\\
	  Harare & Zimbabwe\\
	  Douala & Cameroon\\
	Hargeisa & Republic of Somaliland and Somalia\\
	   Abuja & Nigeria\\
	 Kampala & Uganda\\
	  Bamako & Mali\\
	  Maputo & Mozambique\\
    Antananarivo & Madagascar\\
		 Lusaka & Zambia\\
	 Yaounde & Cameroon\\
     Ouagadougou & Burkina Faso\\
		Conakry & Guinea\\
	  Kumasi & Ghana\\
      Lubumbashi & Democratic Republic of the Congo\\
		  Mbuji & Democratic Republic of the Congo\\
     Brazzaville & Republic of the Congo\\
		   Oran & Algeria\\
	   Benin & Nigeria\\
   Port Harcourt & Nigeria\\
		  Tunis & Tunisia\\
	Freetown & Sierra Leone\\
	 Cotonou & Benin\\
     Vereeniging & South Africa\\
		    Fes & Morocco\\
       Maiduguri & Nigeria\\
	       Monrovia & Liberia\\
	 Port Elizabeth & South Africa\\
		 Huambo & Angola\\
       Ogbomosho & Nigeria\\
		  Zaria & Nigeria\\
	Ndjamena & Chad\\
	\hline
\end{tabular}
\end{center}
\end{table}

\bibliographystyle{pnas}
\bibliography{scibib}